\documentclass[a4paper]{article}

\usepackage{theorem}
\usepackage{latexsym,amssymb,amsfonts,amsmath}
\usepackage{cite}
\usepackage{color}
\usepackage{comment}

\usepackage{graphicx}

\newtheorem{thm}{Theorem}[section]
\newtheorem{lem}[thm]{Lemma}

\newtheorem{pro}[thm]{Proposition}

\theoremheaderfont{\scshape}

\newcommand{\ket}[1]{|#1\rangle}
\newcommand{\bra}[1]{\langle#1|}
\newcommand{\braket}[2]{\langle#1|#2\rangle}

\title{{\Large {\bf Spectral analysis of hierarchical continuous-time quantum walks}}}
\author{
{\small Jir\^{o} Akahori}
\\
{\scriptsize 
Department of Mathematical Sciences, 
College of Science and Engineering, 
Ritsumeikan University, 
Japan
}
\\
{\small Yusuke Ide}
\footnote{To whom correspondence should be addressed. E-mail: ide.yusuke@nihon-u.ac.jp}
, 
{\small Tomoki Kato}
\\
{\scriptsize 
Department of Mathematics, 
College of Humanities and Sciences, 
Nihon University, 
Japan
}
\\
{\small Norio Konno}
\\
{\scriptsize 
Department of Mathematical Sciences, 
College of Science and Engineering, 
Ritsumeikan University, 
Japan
}
\\
{\small Shuhei Mano}
\\
{\scriptsize 
The Institute of Statistical Mathematics, 
Japan
}
\\
{\small Akihiro Narimatsu}
\\
{\scriptsize
Department of Applied Systems and Mathematics, Faculty of Informatics, Kanagawa University, 
Japan
}
}
\date{
}
\begin{document}
\maketitle

\par\noindent
\begin{small}
\par\noindent
{\bf Abstract}
\newline 
In this paper, we introduce hierarchical random walks at first. In this model, we use two types of random walkers, {global and local} walkers. The global walker chooses a local walker at every step, then the chosen local walker moves a single step. After that we construct the corresponding continuous-time quantum walks and discuss its spectral structures. Then we define multi-dimensional continuous-time quantum walk by taking a marginal distribution respect to the global walker.
%\footnote[0]{
%{\it Abbr. title:} Spectral analysis of DTQWs on the path
%}
%\footnote[0]{
%{\it AMS 2000 subject classifications: }
%60F05, 60G50, 82B41, 81Q99
%}
%\footnote[0]{
%{\it PACS: } 
%03.67.Lx, 05.40.Fb, 02.50.Cw
%}
\end{small}

\setcounter{equation}{0}
%%%%%%%%%%%%%%%%%%%%%%%%%%%%%%%%%%%%%%%%%%%%%%%%%%%%%%%
\section{Introduction}\label{intro}
%%%%%%%%%%%%%%%%%%%%%%%%%%%%%%%%%%%%%%%%%%%%%%%%%%%%%%%
In the last 25 years, the theory of quantum walks, quantum counterpart of random walks, has been extensively developed in various fields. It plays fundamental and important roles in both theoretical fields and applications. There are good review articles for these developments such as Kempe\cite{Kempe2003}, Kendon\cite{Kendon2007}, Venegas-Andraca\cite{VAndraca2008,VAndraca2012}, Konno\cite{Konno2008b}, Manouchehri and Wang\cite{ManouchehriWang2013}, and Portugal\cite{Portugal2013}.

In this paper, we propose a method to construct multi-walker version of quantum walks. This is an extension of the previous work \cite{IdeKonnoNarimatsu2025} and also a generalization of the method \cite{DiaconisShahshahani1987} to analyze the Ehrenfest model \cite{EhrenfestEhrenfest1907} by using tensor product of groups. The main idea of proposed method is a hierarchical construction of random walks. In this model, we use two types of random walkers, {global and local} walkers. The global walker chooses a local walker at every step, then the chosen local walker moves a single step. This construction enables us to define the corresponding quantum walks. Although we discuss continuous-time quantum walks in this paper, revealing spectral structures of discrete-time quantum walks can be an interesting future problem. 

The rest of this paper is organized as follows. In Sec. \ref{defHCTQW}, we define our setting of discrete-time random walks, {continuous-time random walks} and continuous-time quantum walks which have hierarchical structures. {After that we discuss the spectral properties of those.} In Sec. \ref{defmCTQW}, we define multi-dimensional continuous-time quantum walks by taking a marginal distribution. {We show a concrete example for the multi-dimensional continuous-time quantum walk which is a generalization of the previous work \cite{IdeKonnoNarimatsu2025}.}
%%%%%%%%%%%%%%%%%%%%%%%%%%%%%%%%%%%%%%%%%%%%%%%%%%%%%%%
\section{Hierarchical continuous-time quantum walk}\label{defHCTQW}
%%%%%%%%%%%%%%%%%%%%%%%%%%%%%%%%%%%%%%%%%%%%%%%%%%%%%%%
Let $G=(V(G),E(G))$ be a finite undirected graph with the vertex set $V(G)$ and the edge set $E(G)$. We consider a discrete-time random walk (DTRW) on $G$ with transition matrix $P_{G}$. We assume existence of the reversible distribution $\{\pi_{G}(j)\neq 0\}_{j\in V(G)}$ with the detailed balance condition. This assumption guarantees symmetrization of $P_{G}$, i.e., if we consider a diagonal matrix $D_{G}=\textrm{diag}\{\pi_{G}(j)\}_{j\in V(G)}$ then the normalized Laplacian matrix 
$
\mathcal{L}_{G}
=
D_{G}^{1/2}\left(I_{\sharp V(G)}-P_{G}\right)D_{G}^{-1/2}
=
I_{\sharp V(G)}-D_{G}^{1/2}P_{G}D_{G}^{-1/2}
$ is symmetric, where $I_{k}$ is the $k$-dimensional identity matrix. 

Now we introduce \textit{hierarchical discrete-time random walk} (hDTRW). 
Let $H=(V(H), E(H))$ be a graph with the vertex set $V(H)=\{0, 1, \ldots , d\}$ and a collection of graphs $(G_{0}, G_{1}, \ldots , G_{d})$ with the vertex sets $V(G_{j})=\{0, 1, \ldots , N_{j}\}$ for each $j=0, 1,\ldots , d$. We write $G=(H;G_{0}, G_{1}, \ldots , G_{d})$ to represent the pair of the graph $H$ and the collection of graphs $(G_{0}, G_{1}, \ldots , G_{d})$. In this model, we consider random walks with its transition matrices $P_{H}, P_{G_{0}}, P_{G_{1}}, \ldots ,P_{G_{d}}$ on the graphs $H, G_{0}, G_{1}, \ldots , G_{d}$. We consider the standard basis $\{\ket{0}, \ket{1}, \ldots , \ket{d}\}$ on $\mathbb{C}^{d+1}$ as column vectors, the transition matrix of the hDTRW $P_{G}$ is defined by
\begin{align*}
P_{G}
&=
\sum_{j=0}^{d}
P_{H}
\ket{j}\bra{j}
\otimes 
\widetilde{P_{G_{j}}}
,
\end{align*}
where $\bra{v}$ is the conjugate transpose of $\ket{v}$, i.e.,  $\bra{v}={}^{\dagger}\ket{v}$ and 
\begin{align*}
\widetilde{A_{G_{j}}}
&=
I_{\sharp V(G_{0})}
\otimes 
\cdots 
\otimes 
I_{\sharp V(G_{j-1})}
\otimes 
A_{G_{j}} 
\otimes 
I_{\sharp V(G_{j+1})}
\otimes 
\cdots 
\otimes 
I_{\sharp V(G_{d})}
,
\end{align*}
for {an arbitrary} matrix $A_{G_{j}}$ on graph $G_{j}$. 

In the hDTRW, there are two types of walkers, \textit{global walker} and \textit{local walker}. The global walker walks on the graph $H$ to choose a graph from $G_{0}, G_{1}, \ldots ,G_{d}$. When the global walker is on a vertex $j\in V(H)$ then the graph $G_{j}$ is chosen and the local walker on $G_{j}$ walks a single step. 

Next we consider the spectral {properties} of $P_{G}$. Because of the assumption, we can consider the spectral decompositions 
$
P_{G_{j}}
=
\sum_{\ell^{(j)}}
\lambda _{\ell^{(j)}}
\left\{
D_{G_{j}}^{-1/2}\ket{v_{\ell^{(j)}}}\bra{v_{\ell^{(j)}}}D_{G_{j}}^{1/2}
\right\}
$ 
by using that of the corresponding normalized Laplacian matrices 
$
\mathcal{L}_{G_{j}}
=
D_{G_{j}}^{1/2}\left(I_{\sharp V(G_{j})}-P_{G_{j}}\right)D_{G_{j}}^{-1/2}
=
\sum_{\ell^{(j)}}
\left(
1-\lambda _{\ell^{(j)}}
\right)
\ket{v_{\ell^{(j)}}}\bra{v_{\ell^{(j)}}}
$
for each $j=0,1,\ldots , d$. 
{Note that the symbol $\ell^{(j)}$ stands for the label of eigenvalues of the matrix $\mathcal{L}_{G_{j}}$. Thus the above summations run on every possible eigenvalues of the matrix $\mathcal{L}_{G_{j}}$.} Let $\ket{v}\in \mathbb{C}^{d+1}$ be an arbitrarily chosen vector. Then we have the action of $P_{G}$ to a vector 
$
\ket{v}\otimes \left( \bigotimes _{j=0}^{d}D_{G_{j}}^{-1/2}\ket{v_{\ell^{(j)}}} \right)
$ 
as follows:
\begin{align*}
P_{G}
\left\{
\ket{v}\otimes \left( \bigotimes _{j=0}^{d}D_{G_{j}}^{-1/2}\ket{v_{\ell^{(j)}}} \right)
\right\}
&=
\left\{
\sum_{j=0}^{d}
P_{H}
\ket{j}\bra{j}
\otimes 
\widetilde{P_{G_{j}}}
\right\}
\left\{
\ket{v}\otimes \left( \bigotimes _{j=0}^{d}D_{G_{j}}^{-1/2}\ket{v_{\ell^{(j)}}} \right)
\right\}
\\
&=
\left\{
P_{H}
\left(
\sum_{j=0}^{d}
\lambda_{\ell^{(j)}}
\ket{j}\bra{j}
\right)
\ket{v}
\right\}
\otimes 
\left( \bigotimes _{j=0}^{d}D_{G_{j}}^{-1/2}\ket{v_{\ell^{(j)}}} \right).
\end{align*}
This shows that when we define diagonal matrix
\begin{align*}
\Lambda ^{(\ell^{(0)}, \ldots , \ell^{(d)})}
=
\sum_{j=0}^{d}
\lambda_{\ell^{(j)}}
\ket{j}\bra{j},
\end{align*}
{for each collection of eigenvalues $\left( \lambda _{\ell^{(0)}},\ldots , \lambda_{\ell^{(d)}} \right)$, }
then the transition matrix $P_{G}$ is represented by
\begin{align*}
P_{G}
=
\sum_{\ell^{(0)}, \ldots , \ell^{(d)}}
P_{H}
\Lambda ^{(\ell^{(0)}, \ldots , \ell^{(d)})}
\otimes 
\left(
\bigotimes _{j=0}^{d}
D_{G_{j}}^{-1/2}\ket{v_{\ell^{(j)}}}\bra{v_{\ell^{(j)}}}D_{G_{j}}^{1/2}
\right).
\end{align*}
{Here the above summation runs on every possible collections of eigenvalues $\left( \lambda _{\ell^{(0)}},\ldots , \lambda_{\ell^{(d)}} \right)$. }
Therefore we obtain eigenvalues and eigenvectors of $P_{G}$ as follows:
\begin{pro}\label{pro:specPG}
If {we obtain the eigenvalue $\lambda _{\ell }^{(\ell^{(0)}, \ldots , \ell^{(d)})}$ and the corresponding eigenvector $\ket{v_{\ell }^{(\ell^{(0)}, \ldots , \ell^{(d)})}}$ the matrix $P_{H}
\Lambda ^{(\ell^{(0)}, \ldots , \ell^{(d)})}$ i.e.}
\begin{align*}
\left\{
P_{H}
\Lambda ^{(\ell^{(0)}, \ldots , \ell^{(d)})}
\right\}
\ket{v_{\ell }^{(\ell^{(0)}, \ldots , \ell^{(d)})}}
=
\lambda _{\ell }^{(\ell^{(0)}, \ldots , \ell^{(d)})}
\ket{v_{\ell }^{(\ell^{(0)}, \ldots , \ell^{(d)})}},
\end{align*}
{for $\ell=0,\ldots ,d$,} then {we have the eigenvalue and the corresponding eigenvectors of $P_{G}$ as }
\begin{align*}
P_{G}
\left\{
\ket{v_{\ell }^{(\ell^{(0)}, \ldots , \ell^{(d)})}}\otimes \left( \bigotimes _{j=0}^{d}D_{G_{j}}^{-1/2}\ket{v_{\ell^{(j)}}} \right)
\right\}
=
\lambda _{\ell }^{(\ell^{(0)}, \ldots , \ell^{(d)})}
\left\{
\ket{v_{\ell }^{(\ell^{(0)}, \ldots , \ell^{(d)})}}\otimes \left( \bigotimes _{j=0}^{d}D_{G_{j}}^{-1/2}\ket{v_{\ell^{(j)}}} \right)
\right\}.
\end{align*}
\end{pro}

%\textcolor{red}{BEGIN}
As we can obtain the eigenvalues and the corresponding eigenvectors for hDTRW by Proposition \ref{pro:specPG}, we can also have that of \textit{hierarchical continuous-time random walk} (hCTRW) befined by
\begin{align*}
P_{G}(t_{0},\ldots ,t_{d})
&=
\sum_{j=0}^{d}
P_{H}
\ket{j}\bra{j}
\otimes 
\widetilde{P_{G_{j}}(t_{j})}
\end{align*}
with 
\begin{align*}
{
\widetilde{P_{G_{j}}(t_{j})}
}
=
\exp\left\{ -t_{j}\left( I_{\sharp V(G_{j})}-P_{G_{j}} \right) \right\}
\end{align*}
for positive constants $t_{j} >0$. For hCTRW, we have more information about its eigenvalues and eigenvectors than hDTRW case. In this case, the action of $P_{G}(t_{0},\ldots ,t_{d})$ to a vector $\ket{v}\otimes \left( \bigotimes _{j=0}^{d}D_{G_{j}}^{-1/2}\ket{v_{\ell^{(j)}}} \right)$ as follows:
\begin{align*}
&P_{G}(t_{0},\ldots ,t_{d})
\left\{
\ket{v}\otimes \left( \bigotimes _{j=0}^{d}D_{G_{j}}^{-1/2}\ket{v_{\ell^{(j)}}} \right)
\right\}
\\
&=
\left[
P_{H}
\left\{
\sum_{j=0}^{d}
\exp\left\{ -t_{j}\left( 1-\lambda _{\ell^{(j)}} \right) \right\}
\ket{j}\bra{j}
\right\}
\ket{v}
\right]
\otimes 
\left( \bigotimes _{j=0}^{d}D_{G_{j}}^{-1/2}\ket{v_{\ell^{(j)}}} \right).
\end{align*}
This shows that when we define diagonal matrices
\begin{align*}
\Lambda ^{(\ell^{(0)}, \ldots , \ell^{(d)})}_{(t_{0}, \ldots , t_{d})}
=
\sum_{j=0}^{d}
\exp\left\{ -t_{j}\left( 1-\lambda _{\ell^{(j)}} \right) \right\}
\ket{j}\bra{j},
\end{align*}
then the transition matrix $P_{G}(t_{0},\ldots ,t_{d})$ is represented by
\begin{align*}
P_{G}(t_{0},\ldots ,t_{d})
=
\sum_{\ell^{(0)}, \ldots , \ell^{(d)}}
P_{H}
\Lambda ^{(\ell^{(0)}, \ldots , \ell^{(d)})}_{(t_{0}, \ldots , t_{d})}
\otimes 
\left(
\bigotimes _{j=0}^{d}
D_{G_{j}}^{-1/2}\ket{v_{\ell^{(j)}}}\bra{v_{\ell^{(j)}}}D_{G_{j}}^{1/2}
\right).
\end{align*}

Focus on the matrices 
$
P_{H}
\Lambda ^{(\ell^{(0)}, \ldots , \ell^{(d)})}_{(t_{0}, \ldots , t_{d})}
$. Because the matrix $\Lambda ^{(\ell^{(0)}, \ldots , \ell^{(d)})}_{(t_{0}, \ldots , t_{d})}$ is positive definite, we can consider its inverse and square root. Noting that the diagonal matrices are always commutative, we have
\begin{align*}
&
P_{H}
\Lambda ^{(\ell^{(0)}, \ldots , \ell^{(d)})}_{(t_{0}, \ldots , t_{d})}
\\
&=
\left(
\Lambda ^{(\ell^{(0)}, \ldots , \ell^{(d)})}_{(t_{0}, \ldots , t_{d})}
\right)^{-1/2}
\left\{
\left(
\Lambda ^{(\ell^{(0)}, \ldots , \ell^{(d)})}_{(t_{0}, \ldots , t_{d})}
\right)^{1/2}
P_{H}
\left(
\Lambda ^{(\ell^{(0)}, \ldots , \ell^{(d)})}_{(t_{0}, \ldots , t_{d})}
\right)^{1/2}
\right\}
\left(
\Lambda ^{(\ell^{(0)}, \ldots , \ell^{(d)})}_{(t_{0}, \ldots , t_{d})}
\right)^{1/2}
\\
&=
\left(
\Lambda ^{(\ell^{(0)}, \ldots , \ell^{(d)})}_{(t_{0}, \ldots , t_{d})}
\right)^{-1/2}
\left\{
\left(
\Lambda ^{(\ell^{(0)}, \ldots , \ell^{(d)})}_{(t_{0}, \ldots , t_{d})}
\right)^{1/2}
D_{H}^{-1/2}
\left(
I_{\sharp V(H)}
-
\mathcal{L}_{H}
\right)
D_{H}^{1/2}
\left(
\Lambda ^{(\ell^{(0)}, \ldots , \ell^{(d)})}_{(t_{0}, \ldots , t_{d})}
\right)^{1/2}
\right\}
\left(
\Lambda ^{(\ell^{(0)}, \ldots , \ell^{(d)})}_{(t_{0}, \ldots , t_{d})}
\right)^{1/2}
\\
&=
\left(
\Lambda ^{(\ell^{(0)}, \ldots , \ell^{(d)})}_{(t_{0}, \ldots , t_{d})}
\right)^{-1/2}
D_{H}^{-1/2}
\left\{
\left(
\Lambda ^{(\ell^{(0)}, \ldots , \ell^{(d)})}_{(t_{0}, \ldots , t_{d})}
\right)^{1/2}
\left(
I_{\sharp V(H)}
-
\mathcal{L}_{H}
\right)
\left(
\Lambda ^{(\ell^{(0)}, \ldots , \ell^{(d)})}_{(t_{0}, \ldots , t_{d})}
\right)^{1/2}
\right\}
D_{H}^{1/2}
\left(
\Lambda ^{(\ell^{(0)}, \ldots , \ell^{(d)})}_{(t_{0}, \ldots , t_{d})}
\right)^{1/2}
\\
&=
\left\{
\left(
\Lambda ^{(\ell^{(0)}, \ldots , \ell^{(d)})}_{(t_{0}, \ldots , t_{d})}
\right)^{-1/2}
D_{H}^{-1/2}
\right\}
\mathcal{L}_{H;(t_{0}, \ldots , t_{d})}^{(\ell^{(0)}, \ldots , \ell^{(d)})}
\left\{
D_{H}^{1/2}
\left(
\Lambda ^{(\ell^{(0)}, \ldots , \ell^{(d)})}_{(t_{0}, \ldots , t_{d})}
\right)^{1/2}
\right\},
\end{align*}
where 
\begin{align*}
\mathcal{L}_{H;(t_{0}, \ldots , t_{d})}^{(\ell^{(0)}, \ldots , \ell^{(d)})}
=
\left(
\Lambda ^{(\ell^{(0)}, \ldots , \ell^{(d)})}_{(t_{0}, \ldots , t_{d})}
\right)^{1/2}
\left(
I_{\sharp V(H)}
-
\mathcal{L}_{H}
\right)
\left(
\Lambda ^{(\ell^{(0)}, \ldots , \ell^{(d)})}_{(t_{0}, \ldots , t_{d})}
\right)^{1/2}.
\end{align*}

Now we define a real symmetric matrix $\mathcal{L}_{G;(t_{0}, \ldots , t_{d})}$ corresponding to the transition matrix $P_{G}(t_{0}, \ldots , t_{d})$, 
\begin{align*}
\mathcal{L}_{G;(t_{0}, \ldots , t_{d})}
=
\sum_{\ell^{(0)}, \ldots , \ell^{(d)}}
\mathcal{L}_{H;(t_{0}, \ldots , t_{d})}^{(\ell^{(0)}, \ldots , \ell^{(d)})}
\otimes 
\left(
\bigotimes _{j=0}^{d}
\ket{v_{\ell^{(j)}}}\bra{v_{\ell^{(j)}}}
\right).
\end{align*}
By definition, we obtain the following spectral decomposition of $\mathcal{L}_{G;(t_{0}, \ldots , t_{d})}$:
\begin{pro}\label{pro:specLG}
The spectral decomposition of $\mathcal{L}_{G;(t_{0}, \ldots , t_{d})}$ is given by
\begin{align*}
\mathcal{L}_{G;(t_{0}, \ldots , t_{d})}
=
\sum_{\ell^{(0)}, \ldots , \ell^{(d)}}
\left[
\sum_{\ell =0}^{d}
\lambda _{\ell ;(t_{0}, \ldots , t_{d})}^{(\ell^{(0)}, \ldots , \ell^{(d)})}
\left\{
\ket{v_{\ell ;(t_{0}, \ldots , t_{d})}^{(\ell^{(0)}, \ldots , \ell^{(d)})}}
\bra{v_{\ell ;(t_{0}, \ldots , t_{d})}^{(\ell^{(0)}, \ldots , \ell^{(d)})}}
\otimes 
\left(
\bigotimes _{j=0}^{d}
\ket{v_{\ell^{(j)}}}\bra{v_{\ell^{(j)}}}
\right)
\right\}
\right], 
\end{align*}
where the spectral decomposition of the matrix $\mathcal{L}_{H;(t_{0}, \ldots , t_{d})}^{(\ell^{(0)}, \ldots , \ell^{(d)})}$ is given by 
\begin{align*}
\mathcal{L}_{H;(t_{0}, \ldots , t_{d})}^{(\ell^{(0)}, \ldots , \ell^{(d)})}
=
\sum_{\ell =0}^{d}
\lambda _{\ell ;(t_{0}, \ldots , t_{d})}^{(\ell^{(0)}, \ldots , \ell^{(d)})}
\ket{v_{\ell ;(t_{0}, \ldots , t_{d})}^{(\ell^{(0)}, \ldots , \ell^{(d)})}}
\bra{v_{\ell ;(t_{0}, \ldots , t_{d})}^{(\ell^{(0)}, \ldots , \ell^{(d)})}}.
\end{align*}
In addition, the spectral decomposition of the deformed matrix $P_{G}(t_{0}, \ldots , t_{d})$ is given by
\begin{align*}
P_{G}(t_{0}, \ldots , t_{d})
=
\sum_{\ell^{(0)}, \ldots , \ell^{(d)}}
\left\{
\sum_{\ell =0}^{d}
\lambda _{\ell ;(t_{0}, \ldots , t_{d})}^{(\ell^{(0)}, \ldots , \ell^{(d)})}
\widetilde{
\ket{v_{\ell ;(t_{0}, \ldots , t_{d})}^{(\ell^{(0)}, \ldots , \ell^{(d)})}}
}
\widetilde{
\bra{w_{\ell ;(t_{0}, \ldots , t_{d})}^{(\ell^{(0)}, \ldots , \ell^{(d)})}}
}
\right\}, 
\end{align*}
where 
\begin{align*}
\widetilde{
\ket{v_{\ell ;(t_{0}, \ldots , t_{d})}^{(\ell^{(0)}, \ldots , \ell^{(d)})}}
}
&=
\left\{
\left(
\Lambda ^{(\ell^{(0)}, \ldots , \ell^{(d)})}_{(t_{0}, \ldots , t_{d})}
\right)^{-1/2}
D_{H}^{-1/2}
\right\}
\ket{v_{\ell ;(t_{0}, \ldots , t_{d})}^{(\ell^{(0)}, \ldots , \ell^{(d)})}}
\otimes 
\left(
\bigotimes _{j=0}^{d}
D_{G_{j}}^{-1/2}\ket{v_{\ell^{(j)}}}
\right),
\\
\widetilde{
\bra{w_{\ell ;(t_{0}, \ldots , t_{d})}^{(\ell^{(0)}, \ldots , \ell^{(d)})}}
}
&=
\bra{v_{\ell ;(t_{0}, \ldots , t_{d})}^{(\ell^{(0)}, \ldots , \ell^{(d)})}}
\left\{
D_{H}^{1/2}
\left(
\Lambda ^{(\ell^{(0)}, \ldots , \ell^{(d)})}_{(t_{0}, \ldots , t_{d})}
\right)^{1/2}
\right\}
\otimes 
\left(
\bigotimes _{j=0}^{d}
\bra{v_{\ell^{(j)}}}D_{G_{j}}^{1/2}
\right).
\end{align*}
\end{pro}
%\textcolor{red}{END}

{Inspired by these arguments}, we define the \textit{hierarchical continuous-time quantum walk} (hCTQW) on $G=(H;G_{0},\ldots ,G_{d})$. Let 
$
U_{H}(t)
=
e^{it\mathcal{H}_{H}}, 
U_{G_{0}}(t)
=
e^{it\mathcal{H}_{G_{0}}}, 
U_{G_{1}}(t)
=
e^{it\mathcal{H}_{G_{1}}}, 
\ldots , 
U_{G_{d}}(t)
=
e^{it\mathcal{H}_{G_{d}}}
$ 
be the time evolution operators of CTQWs on graphs $H, G_{0}, G_{1}, \ldots , G_{d}$, where $i$ is the imaginary unit. If we consider the spectral decomposition of each Hermitian matrix $\mathcal{H}_{G_{j}}\ (j=0,1,\ldots ,d)$ as $\mathcal{H}_{G_{j}}=\sum_{\ell ^{(j)}}\lambda_{\ell ^{(j)}}\ket{v_{\ell^{(j)}}}\bra{v_{\ell^{(j)}}}$ then we can select the minimum eigenvalue $\lambda_{\ell ^{(j)}}^{min}$ and the maximum eigenvalue $\lambda_{\ell ^{(j)}}^{max}$. Note that if we define nonnegative definite matrices 
$
\mathcal{H}_{G_{j}}^{min}
=
\mathcal{H}_{G_{j}}-\lambda_{\ell ^{(j)}}^{min}I_{\sharp V(G_{j})}
=
\sum_{\ell ^{(j)}}
\left(
\lambda_{\ell ^{(j)}}-\lambda_{\ell ^{(j)}}^{min}
\right)
\ket{v_{\ell^{(j)}}}\bra{v_{\ell^{(j)}}}
$ 
and 
$
\mathcal{H}_{G_{j}}^{max}
=
\lambda_{\ell ^{(j)}}^{max}I_{\sharp V(G_{j})}
-
\mathcal{H}_{G_{j}}
=
\sum_{\ell ^{(j)}}
\left(
\lambda_{\ell ^{(j)}}^{max}-\lambda_{\ell ^{(j)}}
\right)
\ket{v_{\ell^{(j)}}}\bra{v_{\ell^{(j)}}}
$ 
then we have the same distributions of CTQWs driven by $e^{it\mathcal{H}_{G_{j}}^{min}}$ and $e^{it\mathcal{H}_{G_{j}}^{max}}$ as $U_{G_{j}}(t)$. 

From now on, we assume that $\mathcal{H}_{G_{j}}\ (j=0,1,\ldots , d)$ are nonnegative definite Hermitian matrices after suitable modifications. Now we define a Hermitian matrix $\mathcal{H}_{G}$ as 
\begin{align*}
\mathcal{H}_{G}
=
\sum_{\ell^{(0)}, \ldots , \ell^{(d)}}
\mathcal{H}_{H}^{(\ell^{(0)}, \ldots , \ell^{(d)})}
\otimes 
\left(
\bigotimes _{j=0}^{d}
\ket{v_{\ell^{(j)}}}\bra{v_{\ell^{(j)}}}
\right)
, 
\end{align*}
where 
\begin{align*}
\mathcal{H}_{H}^{(\ell^{(0)}, \ldots , \ell^{(d)})}
=
\left(
\Lambda ^{(\ell^{(0)}, \ldots , \ell^{(d)})}
\right)^{1/2}
\mathcal{H}_{H}
\left(
\Lambda ^{(\ell^{(0)}, \ldots , \ell^{(d)})}
\right)^{1/2}
\end{align*}
with diagonal matrices 
\begin{align*}
\Lambda ^{(\ell^{(0)}, \ldots , \ell^{(d)})}
=
\sum_{j=0}^{d}
\lambda_{\ell^{(j)}}
\ket{j}\bra{j}
. 
\end{align*}
Because we assume nonnegative definite Hermitian matrices, we can always define 
$
\left(
\Lambda ^{(\ell^{(0)}, \ldots , \ell^{(d)})}
\right)^{1/2}
$. 
The time evolution operator $U_{G}(t)$ for $t\geq0$ is given by
\begin{align*}
U_{G}(t)
=
\exp\left(it\mathcal{H}_{G}\right)
=
\sum_{k=0}^{\infty}\frac{(it)^{k}}{k!}\mathcal{H}_{G}^{k}
. 
\end{align*}
By definition, the spectral decomposition of $U_{G}(t)$ is the following:
\begin{thm}\label{thm:specUG}
The spectral decomposition of $U_{G}(t)$ is given by
\begin{align*}
U_{G}(t)
=
\sum_{\ell^{(0)}, \ldots , \ell^{(d)}}
\left[
\sum_{\ell =0}^{d}
\exp\left(
it
\lambda _{\ell}^{(\ell^{(0)}, \ldots , \ell^{(d)})}
\right)
\left\{
\ket{v_{\ell}^{(\ell^{(0)}, \ldots , \ell^{(d)})}}
\bra{v_{\ell}^{(\ell^{(0)}, \ldots , \ell^{(d)})}}
\otimes 
\left(
\bigotimes _{j=0}^{d}
\ket{v_{\ell^{(j)}}}\bra{v_{\ell^{(j)}}}
\right)
\right\}
\right]
,
\end{align*}
where 
\begin{align*}
\mathcal{H}_{H}^{(\ell^{(0)}, \ldots , \ell^{(d)})}
=
\sum_{\ell =0}^{d}
\lambda _{\ell }^{(\ell^{(0)}, \ldots , \ell^{(d)})}
\ket{v_{\ell }^{(\ell^{(0)}, \ldots , \ell^{(d)})}}
\bra{v_{\ell }^{(\ell^{(0)}, \ldots , \ell^{(d)})}}.
\end{align*}
\end{thm}
%%%%%%%%%%%%%%%%%%%%%%%%%%%%%%%%%%%%%%%%%%%%%%%%%%%%%%%
\section{Multi-dimensional continuous-time quantum walk}\label{defmCTQW}
%%%%%%%%%%%%%%%%%%%%%%%%%%%%%%%%%%%%%%%%%%%%%%%%%%%%%%%
Let $X^{(j)}_{t}\ (j=0,1,\ldots ,d;\ t\geq 0)$ be the random variables of the position of the local walkers on the graph $G_{j}$ at time $t$ with initial state $\ket{\psi_{j}}$. Also we consider the random variable $Y_{t}\ (t\geq 0)$ of the position of the global walker at time $t$ with initial state $\ket{\psi_{H}}$. When we consider an hCTQW on $G=(H;G_{0}, G_{1}, \ldots , G_{d})$, we can obtain marginal distributions of local walkers $X_{t}^{(j)}\ (j=0,1,\ldots d; t\geq 0)$ on $(G_{0}, G_{1}, \ldots , G_{d})$. In this paper, we call the local walkers with marginal distribution as \textit{multi-dimensional continuous-time quantum walk} (mCTQW) induced by the hCTQW. Here, we choose the following marginal distribution:
\begin{align*}
&\mathbb{P}
\left(
X^{(0)}_{t}=k_{0}, X^{(1)}_{t}=k_{1}, \ldots ,X^{(d)}_{t}=k_{d} 
\right)
\\
&=  
\sum_{\ell =0}^{d}
\left|
    \sum_{\ell^{(0)}, \ldots , \ell^{(d)}}
\left\{
\braket{v_{\ell}^{(\ell^{(0)}, \ldots , \ell^{(d)})}}{\psi_{H}}
\exp\left(
it
\lambda _{\ell}^{(\ell^{(0)}, \ldots , \ell^{(d)})}
\right)
\prod _{j=0}^{d}
\braket{k_{j}}{v_{\ell^{(j)}}}\braket{v_{\ell^{(j)}}}{\psi_{j}}
\right\}
\right|^{2}
.
\end{align*}
\begin{comment}
Of course, marginal distributions depend on the choice of 
the initial state $\ket{\psi_{H}}\otimes \left(\bigotimes_{j=0}^{d}\ket{\psi_{j}}\right)$. The simplest choice is
\begin{align*}
\ket{\widetilde{\psi_{H}}}
=
\frac{1}{\sqrt{(d+1)\prod_{j=0}^{d}{N_{j}}}}
\sum_{\ell^{(0)}, \ldots , \ell^{(d)}}
\left\{
\sum_{\ell =0}^{d}
\ket{v_{\ell}^{(\ell^{(0)}, \ldots , \ell^{(d)})}}
\otimes \left(\bigotimes_{j=0}^{d}\ket{\psi_{j}}\right)
\right\}.
\end{align*}
In this case, we have
\begin{align*}
&\widetilde{\mathbb{P}}
\left(
X^{(0)}_{t}=k_{0}, X^{(1)}_{t}=k_{1}, \ldots, X^{(d)}_{t}=k_{d} 
\right)
\\
&=
\frac{1}{(d+1)\prod_{j=0}^{d}{N_{j}}}
\times 
\sum_{\ell =0}^{d}
\left|
    \sum_{\ell^{(0)}, \ldots , \ell^{(d)}}
\left\{
\exp\left(
it
\lambda _{\ell}^{(\ell^{(0)}, \ldots , \ell^{(d)})}
\right)
\prod _{j=0}^{d}
\braket{k_{j}}{v_{\ell^{(j)}}}\braket{v_{\ell^{(j)}}}{\psi_{j}}
\right\}
\right|^{2}
, 
\end{align*}
and call it the \textit{canonical mCTQW}. 
\end{comment}

Here, we consider a special case of mCTQW. Let $H$ be the complete graph with self-loops $\overline{K_{d+1}}$, i.e., $E(\overline{K_{d+1}})=\{(j,k) : 0\leq j\leq k\leq d\}$. We fix a collection of transition probabilities $0 < q_{0}, q_{1}, \ldots , q_{d} < 1$ with $\sum_{j=0}^{d}q_{j} = 1$. Then we define the transition matrix $P_{\overline{K_{d+1}}}$ of a random walk on $\overline{K_{d+1}}$ as 
$
\left(
P_{\overline{K_{d+1}}}
\right)_{j,k}
=q_{k}
$. This means that the transition probability from any vertex $j\in V\left(\overline{K_{d+1}}\right)$ to the vertex $k\in V\left(\overline{K_{d+1}}\right)$ is $q_{k}$ in this model. 

For this model, the reversible distribution is given by 
$
\left\{
\pi_{\overline{K_{d+1}}}(j)=q_{j}
\right\}_{j\in V\left(\overline{K_{d+1}}\right)}.
$ 
Then we have the following Hermitian matrix:
\begin{align*}
\mathcal{H}_{\overline{K_{d+1}}}
=
\left(
\sum_{j=0}^{d}\sqrt{q_{j}}\ket{j}
\right)
\left(
\sum_{j=0}^{d}\sqrt{q_{j}}\bra{j}
\right).
\end{align*}
On the other hand, 
%\textcolor{red}
{
we consider $\mathcal{H}_{G_{j}}=\mathcal{L}_{G_{j}}
=
\sum_{\ell^{(j)}}
\left(
1-\lambda _{\ell^{(j)}}
\right)
\ket{v_{\ell^{(j)}}}\bra{v_{\ell^{(j)}}}
\ (j=0,1, \ldots , d)$ for the local walkers. Note that we use the diagonal matrices as 
$
\Lambda ^{(\ell^{(0)}, \ldots , \ell^{(d)})}
=
\sum_{j=0}^{d}
(1-\lambda_{\ell^{(j)}})
\ket{j}\bra{j}
$
. 
} 
In this case, we obtain 
\begin{align*}
\mathcal{H}_{\overline{K_{d+1}}}^{(\ell^{(0)}, \ldots , \ell^{(d)})}
&=
\left(
\Lambda ^{(\ell^{(0)}, \ldots , \ell^{(d)})}
\right)^{1/2}
\mathcal{H}_{\overline{K_{d+1}}}
\left(
\Lambda ^{(\ell^{(0)}, \ldots , \ell^{(d)})}
\right)^{1/2}
\\
&=
\left\{
\sum_{j=0}^{d}
\sqrt{
\left(
1-\lambda_{\ell^{(j)}}\right
)
q_{j}
}
\ket{j}
\right\}
\left\{
\sum_{j=0}^{d}
\sqrt{
\left(
1-\lambda_{\ell^{(j)}}\right
)
q_{j}
}
\bra{j}
\right\}
.
\end{align*}
If $1-\lambda_{\ell^{(j)}}\neq 0$ for some $j=0,1,\ldots , d$ then we have 
$
\sum_{j=0}^{d}
(
1
-
\lambda_{\ell^{(j)}}
)
q_{j}
\neq 0
$. 
Thus we obtain 
\begin{align*}
\mathcal{H}_{\overline{K_{d+1}}}^{(\ell^{(0)}, \ldots , \ell^{(d)})}
&=
\left\{
\sum_{j=0}^{d}
(
1
-
\lambda_{\ell^{(j)}}
)
q_{j}
\right\}
\times 
\frac{1}{\sum_{j=0}^{d}
(1-\lambda_{\ell^{(j)}})q_{j}}
\left\{
\sum_{j=0}^{d}\sqrt{(1-\lambda_{\ell^{(j)}})q_{j}}\ket{j}
\right\}
\left\{
\sum_{j=0}^{d}\sqrt{(1-\lambda_{\ell^{(j)}})q_{j}}\bra{j}
\right\}
\\
&\quad \quad \quad \quad 
+
0\times
\left[
I_{d+1}
-
\frac{1}{\sum_{j=0}^{d}
(1-\lambda_{\ell^{(j)}})q_{j}}
\left\{
\sum_{j=0}^{d}\sqrt{(1-\lambda_{\ell^{(j)}})q_{j}}\ket{j}
\right\}
\left\{
\sum_{j=0}^{d}\sqrt{(1-\lambda_{\ell^{(j)}})q_{j}}\bra{j}
\right\}
\right].
\end{align*}
{This shows that we obtain eigenvalues and eigenvectors labeled by $\ell $ in Theorem \ref{thm:specUG} for $1-\lambda_{\ell^{(j)}}\neq 0$ for some $j=0,1,\ldots , d$ cases}. 

Otherwise, $\mathcal{H}_{H}^{(\ell^{(0)}, \ldots , \ell^{(d)})}$ is the zero matrix case, we observe the following expression:
\begin{align*}
&\mathcal{H}_{\overline{K_{d+1}}}^{(\ell^{(0)}, \ldots , \ell^{(d)})}
\\
&=
0\times I_{d+1}
\\
&=
\left\{
\sum_{j=0}^{d}
(1-\lambda_{\ell^{(j)}})q_{j}
\right\}
\times 
\left(
\sum_{j=0}^{d}\sqrt{q_{j}}\ket{j}
\right)
\left(
\sum_{j=0}^{d}\sqrt{q_{j}}\bra{j}
\right)
+
0\times
\left\{
I_{d+1}
-
\left(
\sum_{j=0}^{d}\sqrt{q_{j}}\ket{j}
\right)
\left(
\sum_{j=0}^{d}\sqrt{q_{j}}\bra{j}
\right)
\right\}.
\end{align*}
{This shows that we obtain eigenvalues and eigenvectors labeled by $\ell $ in Theorem \ref{thm:specUG} for $1-\lambda_{\ell^{(j)}}= 0$ for all $j=0,1,\ldots , d$ case}. 

Therefore we obtain the following spectral decomposition of the time evolution operator $U_{G}(t)$ of hCTQW:
\begin{lem}\label{lem:hCTQWonK}
The spectral decomposition of $U_{G}(t)$ for $H=\overline{K_{d+1}}$ is given by
\begin{align*}
U_{G}(t)
&=
\sum_{\ell^{(0)}, \ldots , \ell^{(d)}}
\exp\left\{
it
\sum_{j=0}^{d}
(1-\lambda_{\ell^{(j)}})q_{j}
\right\}
\left\{
\ket{v^{(\ell^{(0)}, \ldots , \ell^{(d)})}}
\bra{v^{(\ell^{(0)}, \ldots , \ell^{(d)})}}
\otimes 
\left(
\bigotimes _{j=0}^{d}
\ket{v_{\ell^{(j)}}}\bra{v_{\ell^{(j)}}}
\right)
\right\}
\\
&\quad \quad \quad \quad 
+
\left[
I_{(d+1)\prod_{j=0}^{d}(N_{j}+1)}
-
\sum_{\ell^{(0)}, \ldots , \ell^{(d)}}
\left\{
\ket{v^{(\ell^{(0)}, \ldots , \ell^{(d)})}}
\bra{v^{(\ell^{(0)}, \ldots , \ell^{(d)})}}
\otimes 
\left(
\bigotimes _{j=0}^{d}
\ket{v_{\ell^{(j)}}}\bra{v_{\ell^{(j)}}}
\right)
\right\}
\right]
\\
&=
\sum_{\ell^{(0)}, \ldots , \ell^{(d)}}
\left\{
\ket{v^{(\ell^{(0)}, \ldots , \ell^{(d)})}}
\bra{v^{(\ell^{(0)}, \ldots , \ell^{(d)})}}
\otimes 
\left(
\bigotimes _{j=0}^{d}
\exp\left\{
it
(1-\lambda_{\ell^{(j)}})q_{j}
\right\}
\ket{v_{\ell^{(j)}}}\bra{v_{\ell^{(j)}}}
\right)
\right\}
\\
&\quad \quad \quad \quad 
+
\left[
I_{(d+1)\prod_{j=0}^{d}(N_{j}+1)}
-
\sum_{\ell^{(0)}, \ldots , \ell^{(d)}}
\left\{
\ket{v^{(\ell^{(0)}, \ldots , \ell^{(d)})}}
\bra{v^{(\ell^{(0)}, \ldots , \ell^{(d)})}}
\otimes 
\left(
\bigotimes _{j=0}^{d}
\ket{v_{\ell^{(j)}}}\bra{v_{\ell^{(j)}}}
\right)
\right\}
\right]
,
\end{align*}
where {$N_{j}+1$ be the number of vertices in $G_{j}$ and} 
\begin{align*}
\ket{v^{(\ell^{(0)}, \ldots , \ell^{(d)})}}
=
\begin{cases}
\frac{1}{\sqrt{\sum_{j=0}^{d}
(1-\lambda_{\ell^{(j)}})q_{j}}}
\left(
\sum_{j=0}^{d}\sqrt{(1-\lambda_{\ell^{(j)}})q_{j}}\ket{j}
\right)
\quad &\text{if }1-\lambda_{\ell^{(j)}}\neq 0 \text{ for some $j=0,1,\ldots , d$},
\\
\sum_{j=0}^{d}\sqrt{q_{j}}\ket{j}
\quad &\text{otherwise}.
\end{cases}
\end{align*}
\end{lem}

Using Lemma \ref{lem:hCTQWonK}, we can obtain the marginal distribution as
\begin{align*}
    &\mathbb{P}
    \left(
        X^{(0)}_{t}=k_{0}, X^{(1)}_{t}=k_{1}, \ldots ,X^{(d)}_{t}=k_{d} 
    \right)
    \\
    &=
\left|
\sum_{\ell^{(0)}, \ldots , \ell^{(d)}}
\left\{
\braket{v^{(\ell^{(0)}, \ldots , \ell^{(d)})}}{\psi_{H}}
\prod _{j=0}^{d}
\exp\left(
it
(1-\lambda_{\ell^{(j)}})q_{j}
\right)
\braket{k_{j}}{v_{\ell^{(j)}}}\braket{v_{\ell^{(j)}}}{\psi_{j}}
\right\}
\right|^{2}
\\
&\quad \quad \quad \quad \quad \quad \quad \quad \quad
+
1\times 
\prod_{j=0}^{d}
\left|
\braket{k_{j}}{\psi_{j}}
\right|^{2}
-
\left|
\sum_{\ell^{(0)}, \ldots , \ell^{(d)}}
\left\{
\braket{v^{(\ell^{(0)}, \ldots , \ell^{(d)})}}{\psi_{H}}
\prod _{j=0}^{d}
\braket{k_{j}}{v_{\ell^{(j)}}}\braket{v_{\ell^{(j)}}}{\psi_{j}}
\right\}
\right|^{2}
. 
\end{align*}
If the inner products 
$
\braket{v^{(\ell^{(0)}, \ldots , \ell^{(d)})}}{\psi_{H}}
$ 
are free from the choice of $(\ell^{(0)}, \ldots , \ell^{(d)})$ with a constant $0\leq p\leq 1$ such that $
\left|
\braket{v^{(\ell^{(0)}, \ldots , \ell^{(d)})}}{\psi_{H}}
\right|^{2}
=
p
$, 
then the above mentioned marginal distribution becomes
\begin{align*}
    &\mathbb{P}
    \left(
        X^{(0)}_{t}=k_{0}, X^{(1)}_{t}=k_{1}, \ldots ,X^{(d)}_{t}=k_{d} 
    \right)
    \\
    &=
p
\left|
\sum_{\ell^{(0)}, \ldots , \ell^{(d)}}
\left\{
\prod _{j=0}^{d}
\exp\left(
it
(1-\lambda_{\ell^{(j)}})q_{j}
\right)
\braket{k_{j}}{v_{\ell^{(j)}}}\braket{v_{\ell^{(j)}}}{\psi_{j}}
\right\}
\right|^{2}
+
(1-p)
\left|
\sum_{\ell^{(0)}, \ldots , \ell^{(d)}}
\left\{
\prod _{j=0}^{d}
\braket{k_{j}}{v_{\ell^{(j)}}}\braket{v_{\ell^{(j)}}}{\psi_{j}}
\right\}
\right|^{2}
\\
&=
p
\left|
\prod _{j=0}^{d}
\left\{
\sum_{\ell^{(j)}}
\exp\left(
it
(1-\lambda_{\ell^{(j)}})q_{j}
\right)
\braket{k_{j}}{v_{\ell^{(j)}}}\braket{v_{\ell^{(j)}}}{\psi_{j}}
\right\}
\right|^{2}
+
(1-p)
\left|
\prod _{j=0}^{d}
\left\{
\sum_{\ell^{(j)}}
\braket{k_{j}}{v_{\ell^{(j)}}}\braket{v_{\ell^{(j)}}}{\psi_{j}}
\right\}
\right|^{2}
\\
&=
p
\prod _{j=0}^{d}
\left|
\sum_{\ell^{(j)}}
\exp\left(
it
(1-\lambda_{\ell^{(j)}})q_{j}
\right)
\braket{k_{j}}{v_{\ell^{(j)}}}\braket{v_{\ell^{(j)}}}{\psi_{j}}
\right|^{2}
+
(1-p)
\prod _{j=0}^{d}
\left|
\sum_{\ell^{(j)}}
\braket{k_{j}}{v_{\ell^{(j)}}}\braket{v_{\ell^{(j)}}}{\psi_{j}}
\right|^{2}
\\
&=
p
\prod _{j=0}^{d}
\mathbb{P}
\left(
X_{q_{j}t}^{(j)}=k_{j}
\right)
+
(1-p)
\prod _{j=0}^{d}
\mathbb{P}
\left(
X_{0}^{(j)}=k_{j}
\right)
. 
\end{align*}
We summarize this fact as follows:
\begin{thm}\label{thm:mCTQW}
The distribution of mCTQW for $H=\overline{K_{d+1}}$ with 
$
\mathcal{H}_{\overline{K_{d+1}}}
=
\left(
\sum_{j=0}^{d}\sqrt{q_{j}}\ket{j}
\right)
\left(
\sum_{j=0}^{d}\sqrt{q_{j}}\bra{j}
\right)
$ 
and $\mathcal{H}_{G_{j}}=\mathcal{L}_{G_{j}}\ (j=0,1,\ldots , d)$ is given by
\begin{align*}
    &\mathbb{P}
    \left(
        X^{(0)}_{t}=k_{0}, X^{(1)}_{t}=k_{1}, \ldots ,X^{(d)}_{t}=k_{d} 
    \right)
    \\
    &=
\left|
\sum_{\ell^{(0)}, \ldots , \ell^{(d)}}
\left\{
\braket{v^{(\ell^{(0)}, \ldots , \ell^{(d)})}}{\psi_{H}}
\prod _{j=0}^{d}
\exp\left(
it
(1-\lambda_{\ell^{(j)}})q_{j}
\right)
\braket{k_{j}}{v_{\ell^{(j)}}}\braket{v_{\ell^{(j)}}}{\psi_{j}}
\right\}
\right|^{2}
\\
&\quad \quad \quad \quad \quad \quad \quad \quad \quad
+
1\times 
\prod_{j=0}^{d}
\left|
\braket{k_{j}}{\psi_{j}}
\right|^{2}
-
\left|
\sum_{\ell^{(0)}, \ldots , \ell^{(d)}}
\left\{
\braket{v^{(\ell^{(0)}, \ldots , \ell^{(d)})}}{\psi_{H}}
\prod _{j=0}^{d}
\braket{k_{j}}{v_{\ell^{(j)}}}\braket{v_{\ell^{(j)}}}{\psi_{j}}
\right\}
\right|^{2}
. 
\end{align*}

If the inner products 
$
\braket{v^{(\ell^{(0)}, \ldots , \ell^{(d)})}}{\psi_{H}}
$ 
are free from the choice of $(\ell^{(0)}, \ldots , \ell^{(d)})$ with a constant $0\leq p\leq 1$ such that $
\left|
\braket{v^{(\ell^{(0)}, \ldots , \ell^{(d)})}}{\psi_{H}}
\right|^{2}
=
p
$, 
then the distribution becomes
\begin{align*}
    &\mathbb{P}
    \left(
        X^{(0)}_{t}=k_{0}, X^{(1)}_{t}=k_{1}, \ldots ,X^{(d)}_{t}=k_{d} 
    \right)
    =
p
\prod _{j=0}^{d}
\mathbb{P}
\left(
X_{q_{j}t}^{(j)}=k_{j}
\right)
+
(1-p)
\prod _{j=0}^{d}
\mathbb{P}
\left(
X_{0}^{(j)}=k_{j}
\right)
. 
\end{align*}
\end{thm}

Note that $p=1$ case is appeared in the previous work \cite{IdeKonnoNarimatsu2025}. Therefore this model is an extension of it. 
%%%%%%%%%%%%%%%%%%%%%%%%%%%%%%%%%%%%%%%%%%%%%%%%%%%%%%%
\par
\
\par\noindent
{\bf Acknowledgments.} 
A. N. is partially supported by the Grant-in-Aid for Young Scientists of Japan Society for the Promotion of Science (Grant No. JP23K13017). 
%The author thank the anonymous referee for useful comments to increase the quality of the manuscript. 
%\par
%\
%\par
%%%%%%%%%%%%%%%%%%%%%%%%%%%%%%%%%%%%%%%%%%%%%%%%%%%%%%%

\begin{small}

\end{small}

\end{document}